\documentstyle[fleqn,twoside]{article}

\makeatletter

\def\noi{\noindent}

\renewcommand{\section}{\@startsection{section}{1}{0pt}%
        {-3.5ex plus -1ex minus -.2ex}{2.3ex plus .2ex}%
        {\large\bf\protect\raggedright}}

\renewcommand{\subsection}{\@startsection{subsection}{2}{0pt}%
        {-3ex plus -1ex minus -.2ex}{1.4ex plus .2ex}%
        {\normalsize\bf\protect\raggedright}}

\renewcommand{\thesubsubsection}%
        {\arabic{section}.\arabic{subsection}.\arabic{subsubsection}.}

\renewcommand{\@oddhead}{\raisebox{0pt}[\headheight][0pt]{%
   \vbox{\hbox to\textwidth{\rightmark \hfil \rm \thepage \strut}\hrule}}}
\renewcommand{\@evenhead}{\raisebox{0pt}[\headheight][0pt]{%
   \vbox{\hbox to\textwidth{\thepage \hfil \leftmark \strut}\hrule}}}
\newcommand{\heads}[2]{\markboth{\protect\small\it #1}{\protect\small\it #2}}
\newcommand{\Acknow}[1]{\subsection*{Acknowledgement} #1}

\newcommand{\Arthead}[3]{ \setcounter{page}{#2}\thispagestyle{empty}\noi
    \unitlength=1pt \begin{picture}(500,40)
        \put(0,58){\shortstack[l]{\small\it Gravitation \& Cosmology,
                        \small\rm Vol. 5 (1999), No. #1, pp. #2--#3\\
        \footnotesize \copyright \ 1999 \ Russian Gravitational Society} }
    \end{picture}          }
\newcommand{\Title}[1]{\noi {\Large #1} \\}

\newcommand{\Authors}[4]{\noi
        {\large\bf #1\dag\ #2\ddag}\medskip\begin{description}
        \item[\dag]{\it #3} \item[\ddag]{\it #4}\end{description}}
\newcommand{\Rec}[1]{\noi {\it Received #1} \\}

\newcommand{\Abstract}[1]{\vskip 2mm \begin{center}
        \parbox{16.4cm}{\small\noi #1} \end{center}\medskip}

\newcommand{\foom}[1]{\protect\footnotemark[#1]}

\newcommand{\email}[2]{\footnotetext [#1]{e-mail: #2}
		\addtocounter{footnote}{#1} }

\newcommand{\Ref}[1]{Ref.\,\cite{#1}}
\newcommand{\sect}[1]{Sec.\,#1}
\def\nq{\hspace*{-1em}}
\def\nqq{\hspace*{-2em}}
\def\nhq{\hspace*{-0.5em}}

\def\cm{\hspace*{1cm}}
\def\inch{\hspace*{1in}}
\def\wide{\mbox{$\dst\vphantom{\int}$}}

\def\ten#1{\mbox{$\,\cdot\, 10^{#1}$}}
\def\deg{\mbox{${}^\circ$}\ }                     

\def\al{&\nhq}
\def\lal{&&\nqq {}}
\def\eq{Eq.\,}
\def\eqs{Eqs.\,}
\def\beq{\begin{equation}}
\def\eeq{\end{equation}}
\def\bear{\begin{eqnarray}}
\def\bearr{\begin{eqnarray} \lal}
\def\ear{\end{eqnarray}}
\def\earn{\nonumber \end{eqnarray}}
\def\nn{\nonumber\\ {}}

\def\nnn{\nonumber\\ \lal }
\def\nnnv{\nonumber\\[5pt] \lal }
\def\yy{\\[5pt] {}}
\def\yyy{\\[5pt] \lal }
\def\eql{\al =\al}

\def\dst{\displaystyle}
\def\tst{\textstyle}

\def\fract#1#2{{\tst\frac{#1}{#2}}}

\def\half{{\fract{1}{2}}}

\def\e{{\,\rm e}}

\makeatother

\newcommand{\Picture}[4]{
	\begin{figure} 	\centering \unitlength=1mm
	\begin{picture}(82.5,#2)
		\put(0,0){\line(0,1){#2}}            
		\put(0,0){\line(1,0){82.5}}
		\put(82.5,0){\line(0,1){#2}}
		\put(0,#2){\line(1,0){82.5}}
	\put(#1,#2){#3}                       \end{picture}
        \caption{\protect\small #4}  \medskip \hrule \end{figure} }

\def\Gun{\mbox{${\rm N}\cdot{\rm m}^2/{\rm kg}^2$}}
\def\flun{\mbox{${\rm cm}^{-2}{\rm s}^{-1}$}\ }  
\def\qg{\mbox{${\rm CGSE}_q$}}

\def\ME{M_{\oplus}}
\def\RE{R_{\oplus}}
\def\Horb{H_{\rm orb}}

\def\rt{\vec r{}^{\rm \ th}}
\def\xt{x{}^{\rm th}}
\def\yt{y{}^{\rm th}}

\def\gsim{\mathrel{
    \raisebox{3pt}{$\mathop{>}\limits_{\displaystyle \sim}$}
		  }}
\def\lsim{\mathrel{
   \raisebox{3pt}{$\mathop{<}\limits_{\displaystyle \sim}$}
		  }}
\def\z{\mbox{$\phantom{0}$}}
\def\zz{\mbox{$\phantom{.0}$}}
\def\gapup{\vphantom{\raisebox{1.5mm}{0}}}

\heads
{A.D. Alexeev, K.A. Bronnikov, N.I. Kolosnitsyn,
      M.Yu. Konstantinov, V.N. Melnikov and A.J. Sanders}
      {SEE Project: Current Status and New Estimates}

\begin{document}
\twocolumn[
\Arthead{1 (17)}{67}{78}

\Title{SEE PROJECT FOR TESTING GRAVITY IN SPACE:\yy
       CURRENT STATUS AND NEW ESTIMATES}

\Authors{A.D. Alexeev, K.A. Bronnikov, N.I. Kolosnitsyn,
	    M.Yu. Konstantinov, \\ V.N. Melnikov\foom 1}
         {and A.J. Sanders\foom 2}
         {RGS, 3-1 M. Ulyanovoy St., Moscow 117313, Russia}
         {Dept. of Physics and Astronomy, University of Tennessee,
	     Knoxville, TN 37996-1200, USA}

\Rec{20 December 1998}

\Abstract
    {We describe some new estimates concerning the recently proposed SEE
    (Satellite Energy Exchange) experiment for measuring the gravitational
    interaction parameters in space. The experiment entails precision
    tracking of the relative motion of two test bodies (a heavy ``Shepherd",
    and a light ``Particle") on board a drag-free capsule. The new estimates
    include (i) the sensitivity of Particle trajectories and $G$ measurement
    to the Shepherd quadrupole moment uncertainties; (ii) the measurement
    errors of $G$ and the strength of a putative Yukawa-type force whose
    range parameter $\lambda$ may be either of the order of a few metres or
    close to the Earth radius; (iii) a possible effect of the Van Allen
    radiation belts on the SEE experiment due to test body electric
    charging.  }

]  
\email 1 {melnikov@rgs.phys.msu.su}
\email 2 {asanders@utkux.utcc.utk.edu}

\section {Introduction}

    The SEE (Satellite Energy Exchange) concept of a space-based
    gravitational experiment was suggested in the early 90s \cite{SD92}
    and was aimed at precisely measuring the gravitational interaction
    parameters: the gravitational constant $G$, possible
    violations of the equivalence principle measured by the
    E\"otv\"os parameter $\eta$, time variations of $G$, and
    hypothetical non-Newtonian gravitational forces
    (parametrized by the Yukawa strength $\alpha$ and range
    $\lambda$). Such tests are intended to fill gaps left by current
    methods of ground-based experimentation and observation of astronomical
    phenomena. The significance of new measurements is quite evident
    since nearly all modified theories of gravity and unified
    theories predict some violations of the Equivalence Principle
    (EP), either by deviations from the Newtonian law
    (inverse-square-law, ISL) or by composition-dependent (CD) gravity
    accelerations, due to the appearance of new possible massive particles
    (partners); time variations of $G$ are also generally predicted
    \cite{dSMP, Mel94}.

    The idea of the SEE method is to study the relative motion of two
    bodies on board a drag-free Earth satellite using horseshoe-type
    trajectories, previously known in planetary satellite astronomy:
    if the lighter body (the "Particle") is moving along a lower
    orbit than the heavier one (the "Shepherd") and approaching
    from behind, then the Particle almost overtakes the Shepherd,
    but it gains energy due to their
    gravitational interaction, passes therefore to a higher orbit and
    begins to lag behind. The interaction phase can be studied within a
    drag-free capsule (a cylinder up to 20 m long, about 1 m in diameter)
    where the Particle can loiter as long as $10^5$ seconds.
    It was claimed that the SEE method exceeded in accuracy all other
    suggestions, at least with respect to $G$ and $\alpha$ for $\lambda$ of
    the order of metres. Some design features were considered, making it
    possible to reduce various sources of error to a negligible level.
    It was concluded, in particular, that the most favourable orbits are
    the sun-synchronous, continuous sunlight orbits situated at altitudes
    between 1390 and 3330 km.

    Since the origination of the SEE concept, the development
    has focused on critical analyses and critical hardware requirements.
    All indications from this work are that the SEE concept is feasible and
    practicable \cite{iztech}.  A ``blue ribbon" Theory Advisory Group,
    formed two years ago to critique Project-SEE activities and goals, has
    concluded that they are sound.

    This paper presents some new evaluations concerning the opportunities of
    the SEE concept and its yet-unresolved difficulties. In \sect 2, for
    comparison, we briefly outline the current status of terrestrial and
    astronomical determination of the gravitational interaction parameters
    to be measured by the SEE method. In \sect 3, on the basis of computer
    simulations of Particle trajectories, we estimate the requirements for
    the Shepherd quadrupole moment uncertainty. \sect 4 shows the results of
    simulations of the measurement procedure itself, which enables us to
    estimate the possible measurement accuracy with respect to $G$ and
    $\alpha$ for $\lambda$ of the order of either metres or the Earth's
    radius.  \sect 5 discusses a spurious effect of test body electric
    charging when the satellite orbit passes through the Van Allen radiation
    belts, rich in high-energy protons. \sect 6 is a conclusion.

    In what follows, the term ``orbit" applies to satellite (or
    Shepherd) motion around the Earth, while the words ``trajectory" or
    ``path" apply to Particle motion with respect to the Shepherd inside the
    capsule.

\section{State of the art: a brief survey}

    Since gravitational forces are so very small, precision-measurement
    techniques have been at the core of terrestrial gravity research for two
    centuries. However, evidence is increasingly accumulating which
    indicates that terrestrial methods have plateaued in accuracy and are
    unlikely to achieve significant accuracy gains in the future
    \cite{Gi97}.  For example, the uncertainty in the gravitational constant
    $G$ had been accepted as 128 ppm for nearly two decades, and the actual
    uncertainty in $G$ --- as indicated by the scatter of results among
    recent experiments which claim high accuracy --- is roughly the same
    (about 140 ppm). We discuss below the situation with respect to several
    key measurements.

\subsection{Terrestrial determinations of $G$}

    The Luther \& Towler determination of $G$ in 1982 \cite{LuTh82},
    with the result $(6.6726\pm 0.0005) \ten{-11}$ \Gun\ and other,
    less precise experiments gave rise to the current official CODATA value
    of $G$, viz. 6.67259\ten{-11} \Gun\ with an error of 128
    ppm. Several still other experiments which also claimed high precision
    were ignored by CODATA because of inadequate documentation of systematic
    errors.

    There is considerable evidence that the uncertainty in $G$ has
    plateaued at about 100 ppm. At a recent (November 23--24, 1998)
    conference in
    London, several new (1998) determinations of $G$ were reported.  The
    obtained values for $G$ (in units of $10^{11}$ \Gun) and their estimated
    error $\delta G/G$ in ppm are as follows:
\begin{center}
\begin{tabular}{llr}
    Fitzgerald and Armstrong		      &	6.6742	&   90
							      \\
    \cm (New Zealand) \cite{FiArm}	      &	6.6746	&  134
							      \yy
    Nolting et al. (Zurich) \cite{Nolting}    &	6.6749	&  210
							      \yy
    Meyer et al. (Wuppethal) \cite{Meyer}     &	6.6735	&  240
							      \yy
    Karagioz et al. (Moscow) \cite{Karag}     &	6.6729	&   75
							      \yy
    CODATA (1986)	          		      &	6.67259	&  128
\end{tabular}
\end{center}

    Obviously, most of the stated errors are of the order 100 ppm.  Moreover,
    the scatter (1-sigma) about the mean is about 140 ppm.  Some of the
    investigators still hope for accuracy of 10 ppm.  It remains to be seen
    whether they will be able to report error estimates of this size or,
    more importantly, whether their respective values for $G$ will actually
    agree within 10 ppm.

    We note that this analysis is perhaps unduly optimistic since it
    excludes one extremely bad outlier: the very careful and well documented
    experiment by the Physikalisch-Technische Bundesanstalt in the late
    1980s and early 1990s, which obtained a series of values for $G$ that
    were consistenly above the results of most experimenters by about 6000
    ppm (0.6\,\% !), while claiming an error of about 100 ppm \cite{PTB}. No
    explanation for such a large discrepancy has been found.

    It might seem that the problems of terrestrial apparatus must inexorably
    yield to new technologies --- that the promise of ever increasing
    sensitivities would also lead to ever improving accuracy.  However, this
    may not be true, since it is various systematic errors which limit the
    ultimate attainable accuracy in terrestrial experiments \cite{Gi97}.

\subsection{Terrestrial tests of the equivalence principle (EP) and search
    for Yukawa forces}

    The EP may be tested by searching for either violations of the
    inverse-square law (ISL) or composition-dependent (CD) effects in
    gravitational free fall.

    In the watershed year of 1986, Fischbach startled the physics
    community by showing that E\"otv\"os's famous turn-of-the-century
    experiment is much less decisive as a null result than was generally
    believed \cite{Fis86}. Prior to this time, experiments by
    Dicke \cite{Roll64} and Braginsky \cite{Brag71} had demonstrated the
    universality of free fall (UFF) to very high accuracy with respect to
    several metals falling in the gravitational field of the Sun
    (the E\"otv\"os parameter $\eta$ was ultimately found to be smaller than
    $10^{-12}$).  The interpretation of these results at the time was that
    they validated UFF.

    It was implicit that any violation would have infinite range, like
    gravity \cite{Adel94}. During the 1970s and early 1980s there was also a
    flurry of activity concerning possible ISL violations, which eventually
    led to null results at the levels of precision then available (Fujii,
    \cite{Fuj71-72}, Long \cite{Long76-84}).

    Since 1986 it has become customary to parametrize possible apparent
    EP violations as if due to a Yukawa particle with a Compton wavelength
    $\lambda$. This approach unites both ISL and CD effects very naturally,
    while the parameter values in the Yukawa potential suggest which
    experimental conditions are required to detect the new interaction.

    Following Fischbach's conjecture, ISL and CD tests were
    undertaken by many investigators. Although a number of anomalies
    were initially reported, nearly all of these were eventually explained in
    terms of overlooked systematic errors or extreme sensitivity to models,
    while most investigators obtained null results. By far the tightest
    bounds are those obtained by Adelberger and his ``Eot-Wash" group at
    the University of Washington \cite{Adel+}.
    This group expects a further improvement of at least an order of
    magnitude \cite{Adel97}.
    A positive result for a deviation from the Newtonian law (ISL) was
    obtained (and interpreted in terms of a Yukawa-type potential) in the
    range of 20 to 500 m by Achilli and colleagues \cite{Ach}; this needs to
    be verified in other independent experiments.

    For reviews of terrestrial searches for non-Newtonian gravity, see
    \cite{Adel94, FiG, Franklin}. The opportunities of the SEE concept in
    this respect are discussed in Refs.\,\cite{SD92, iztech} and in the
    present paper.

    The UFF is still in the scope of the current experimental
    projects, and the SEE concept suggests here a progress of 3
    to 4 orders of magnitude as compared with \Ref{Brag71}. Only one
    project, STEP (Satellite Test of the EP) promises a greater
    progress but meets some significant problems of its own \cite{STEP},
    connected, in particular, with the radiation belts.

\section
{Simulations of Particle trajectories and the Shepherd quadrupole moment}

In the previous studies of the SEE project it was assumed that the capsule
was about 20 m long and the initial Shepherd-Particle separation $x_0$
along the capsule axis was as
great as 18 m; some estimations were also made for $5$ m $\leq x_0\leq
10$ m. The Shepherd mass was taken to be $M=500$ kg and the Particle mass
$m=0.1$ kg. The present study retains these values.

In what follows we describe some characteristic features of Particle
trajectories with a goal to determine
their sensitivity to the uncertainty of the Shepherd quadrupole
moment $J_2$ for $x_0\geq 5$ m.
As in our previous studies, the capsule diameter is supposed to be 1 m.

The reason for considering the quadrupole moment uncertainty is
technological by origin. Namely, it is hard to produce a spherically
symmetric Shepherd to a required accuracy and, instead, it has been
suggested \cite{SD92} to use a Cook-Marussi stack of cylinders with $J_2
>0$, which may be manufactured more easily. A slow rotation of the Shepherd
with $J_2>0$ will stabilize its position and orientation.

The value of $J_2$ can be provided with some uncertainty $\delta J_2$.
To avoid the inclusion of $\delta J_2$ in the set of parameters to be
determined in the experiment, it is useful to know which values of $\delta
J_2$ will be negligible, since the growth of the number of parameters
leads to serious problems in data processing.

\subsection{Equations of motion and the initial data}

Assuming that the relative motion of the test bodies inside the capsule
occurs in the satellite orbital plane,
the reduced Lagrangian of the Particle motion reads
\bearr
\label{lagr}
    L=\frac M2(\dot R^2+R^2\dot \varphi ^2)                        \nnn
   \cm +\frac m2\left[ \dot r^2+r^2(\dot \varphi +\dot \psi )^2\right]
        			+G\frac{\ME  m}r                 \nnn
\nq\ +G\frac{Mm}s\left\{ 1+J_2\left(\frac{r_s}s\right)^2\!
      P_2(\cos\theta )\right\}
      			\left( 1+\alpha e^{-s/\lambda }\right)
\ear
where $( R,\varphi)$ are the Earth-centred polar coordinates of the
Shepherd in the orbital plane;
$r=\sqrt{(R{+}y)^2{+}x^2}$ and $\psi$
are the Earth-centred polar coordinates of the Particle;
$x$ and $y$ are the Shepherd-centred Particle coordinates, where $x$ is the
``horizontal" one, i.e., along the orbit and simultaneously along the
capsule and $y$ is the ``vertical" one, along the Earth-Shepherd radius
vector;
$s=\sqrt{x^2+y^2}$ is the Particle-Shepherd separation;
$\ME $, $M$ and $m$ are the Earth, Shepherd and Particle masses,
respectively;
$J_2$ is the quadrupole moment of the Shepherd, $r_s$ is its
radius and $P_2$ is the Legendre polynomial
\[
	P_2(\cos \theta )=\frac{3\cos ^2\theta -1}2,
\]
where $\theta$ is the angle between the line connecting the centres of
the test bodies and the Shepherd equatorial plane. It is easy to see
that if the Shepherd symmetry axis is in its orbital plane, then
$\theta =\theta _0=-\arctan (y/x) +\varphi $. If the symmetry
axis of Shepherd is orthogonal to its orbital plane, then $\theta =0$. In
general, if $\chi$ is the angle between the Shepherd symmetry axis
and its orbital plane, then $\theta =\theta _0\cos \chi $. Hence the
influence of $J_2$ on the Particle motion is
minimum if the Shepherd symmetry axis lies in its orbital plane
and is maximum if they are mutually orthogonal.

For simplicity (and taking into account the corresponding estimate)
we neglect the influence of the Particle on the Shepherd, so the
Shepherd trajectory is considered to be given. Then, varying
the above Lagrangian with respect to $x$ and $y$, taking into account that
$M\gg m$ and $R\gg s$, we arrive at the following equations of Particle
motion with respect to the Shepherd:
\def\M{\overline{M}}
\bearr
\frac{d^2x}{dt^2}=2\dot y\dot \varphi +x\left\{ \dot \varphi ^2-\frac{G\ME}{
r^3}\right\} -\frac{2\dot R\dot \varphi y}R
			 \nnn
-\frac{G\M}{s^3}x\left\{
1+J_2\left( \frac{r_s}s\right) ^2\! P_2(\cos \theta )\right\}
			 \nnn
-\alpha x\frac{G\M}{s^2}\left\{ 1+J_2\left( \frac{r_s}s\right) ^2\! P_2(\cos
\theta )\right\}\! \left(\frac 1s+\frac 1\lambda \right) \e^{-s/\lambda}
			 \nnn                                  \label{eqnx}
-\frac{G\M r_0^2}{2s^5}J_2\left( 1+\alpha e^{-s/\lambda }\right) \times
			 \nnn \cm
\times \left[ x(1+3\cos 2\theta )+3y\sin 2\theta \cos \chi \right]
				\yyy
\frac{d^2y}{dt^2}=-2\dot x\dot \varphi +(R+y)\left\{ \dot \varphi ^2-\frac{
G\ME }{r^3}\right\} +\frac{2\dot R\dot \varphi x}R
			 \nnn
-\frac{G\M}{s^3}y\left\{
1+J_2\left( \frac{r_s}s\right) ^2\! P_2(\cos \theta )\right\}
			 \nnn
-\alpha y\frac{G\M}{s^2}\left( \frac 1s+\frac 1\lambda \right) \!\left\{
1+J_2\left( \frac{r_s}s\right) ^2\! P_2(\cos \theta )\right\} e^{-s/\lambda}
			 \nnn				       \label{eqny}
+\frac{G\M r_0^2}{s^5}J_2\left( 1+\alpha e^{-s/\lambda }\right) \times
			 \nnn \cm
\times  \left[ 3x\sin 2\theta \cos \chi +y(1-3\cos \theta )\right]
\ear
where $\M = M+m$.

Two kinds of initial conditions for \eqs (\ref{eqnx}) and (\ref{eqny})
were used during the simulations. First, we used the so-called ``standard''
initial conditions, taking
the Particle velocity components $\dot x(0)$ and $\dot y(0)$ corresponding
to its unperturbed (i.e., without the $M-m$ interaction) orbital motion
distinguished from the Shepherd's orbit only by its radius (for circular
orbits) or major semiaxis (for elliptic orbits). Assuming that the Particle
motion begins right at the moment when the Shepherd passes its perigee,
these conditions have the form
\bearr
	\label{initcond}
   x(0)=x_0, \cm\cm  y(0)=y_0,\nnn
  \dot x(0)=\frac{\omega
	e'y_0}{2(1-e)^2},\cm  \dot y(0)=-\frac{\omega ex_0}{e'(1-e)}
\ear
where $\omega ^2=G\ME /R_0^3$, $R_0$ is the Shepherd orbital
radius (at the perigee), $e$ is the orbital eccentricity and $e'=\sqrt{1-e^2}$.

For clearness, the relations (\ref{initcond}) are written in the linear
approximation in the variables $x$ and $y$. Higher-order approximations were
used in the simulation process as well.

The second kind of initial conditions correspond to small variations of
initial velocities with respect to their ``standard'' values.

The set of equations (\ref{eqnx})--(\ref{eqny}) was solved numerically using
the software developed previously \cite{iztech} to analyze the SEE project.

On the basis of numerical solution of \eqs (\ref{eqnx}) and (\ref{eqny}), we
considered two types of Particle trajectories, corresponding to different
choices of the initial data: (i) approximately U-shaped ones and (ii)
cycloidal ones, containing loops (see more details on the trajectories in
\cite{SD92,iztech}), for orbital altitudes $\Horb=500$, 1500 and 3000 km.
The uncertainty $\delta J_2$ ranged in the interval $10^{-3}\div 10^{-5}$;
and initial Particle position changed in the range $6$ m $\leq x_0\leq 18$
m, $-25$ cm $\leq y_0\leq -5$ cm.

For $\Horb=500$ km, all U-shaped paths contained a sinusoidal component
(with the orbital frequency), starting at $x_0\leq 8$ m, while
for $ \Horb=1500$ and $3000$ km it was present in paths
starting at $ x_0\leq 10$ m. In other families of U-shaped trajectories
a sinusoidal component was present only in the case $|y_0|\geq 20$ cm.

\subsection{Restrictions on the Shepherd quadrupole moment uncertainty}

Small Shepherd quadrupole moment uncertainties $\delta J_2$ create
small displacements $\delta\vec r$ of a Particle trajectory
with respect to unperturbed one, $\vec r_0(t)$:
\[
\delta \vec{r}= \vec r_j(t)-\vec r_0(t)
\]
where $\vec r_j$ is the perturbed path.
Instead of the full displacement $\delta \vec r$, a
displacement $\delta x$ along the $x$ axis may be considered
since, by numerical simulations, displacements along the $y$ axis
are an order of magnitude smaller than $\delta x$.

Numerical simulations show that in the whole range of the above initial
conditions the displacement $\delta x$ is (as it should naturally be) a
linear function of $\delta J_2$; for $\delta J_2=10^{-4}$.
For the case when the Shepherd symmetry axis is located in the orbital
plane, the maximum values of $\delta x$ for U-shaped trajectories are given
in Table 1.

\begin{table}
\noi
{\bf Table 1.}\
Displacements of U-shaped trajectories under $\delta J_2=10^{-4}$ for
the Shepherd symmetry axis in its orbital plane.
The second line shows $x_0$.

\begin{center}
\begin{tabular}{|r|r|r|r|r|r|}
\hline       \wide
   $y_0$ & \multicolumn{5}{|c|}{$\wide \delta x_{\max} \times 10^{7}$ (m)}\\
\cline{2-6}  \wide
 (cm)      & 18 m     & 10 m  & 8 m    &  6 m & 4 m   \\
\hline       \gapup
       -25 & 13.3\z    & 7.52  & 7.98   & 8.57  & 14.90 \\
       -20 & 5.51      & 4.65  & 4.83   & 5.51  & 5.84  \\
       -15 & 1.98      & 2.87  & 3.03   & 3.3\z & 2.18  \\
       -10 & 0.56      & 1.59  & 1.7\z  & 1.88  & 0.66  \\
       - 5 & 0.11      & 0.59  & 0.62   & 0.69  & 0.14  \\
\hline
\end{tabular}
\end{center}
\end{table}

One can conclude that, if the distance measurement error is $10^{-6}$ m,
for most of the trajectories the uncertainty $\delta J_2 = 10^{-4}$ is
admissible.

The increase of $\delta x$ for small values of $|y_0|$ is explained
by a large displacement of the turning point towards the Shepherd.

When the Shepherd symmetry axis is orthogonal to the orbital plane, these
estimations change as shown in Table 2.

\begin{table}
\noi {\bf Table 2.}
Displacements of U-shaped trajectories under $\delta J_2=10^{-4}$ for
the Shepherd symmetry axis orthogonal to its orbital plane.
The second line shows $x_0$.

\begin{center}
\begin{tabular}{|r|r|r|r|r|r|}
\hline            \wide
   $y_0$ & \multicolumn{5}{|c|}{$\wide \delta x_{\max} \times 10^{7}$ (m)}\\
\cline{2-6}       \wide
   (cm)     & 18 m  & 10 m  & 8 m    & 6 m    & 4 m   \\
\hline            \gapup
       -25 & 54.2\z  & 62.4\z  & 68.9\z   & 81\z\z\,  & 57\z\z\,\\
       -20 & 21.9\z  & 25.8\z  & 28.1\z   & 32.2\z    & 23.3\z  \\
       -15 &  7.94   & 10.2\z  & 11.5\z   & 14\z\z\,  &  8.7\z  \\
       -10 &  2.25   & 3.42    & 4.15     &  5.45     &  2.65  \\
       - 5 &  0.45   & 0.83    & 1.05     &  1.47     &  0.58  \\
\hline
\end{tabular}
\end{center}
\end{table}

For cycloidal trajectories these values are approximately an order
of magnitude smaller than those for the U-shaped ones.
%
%

It was also found that $\delta x$ decreases with increasing
orbital altitude $\Horb$. Table 3 shows, as an example, the displacements of
U-shaped trajectories with $x_0=18$ m for $\Horb=500$, 1500 and 3000 km
and the Shepherd symmetry axis located in its orbital plane.

\begin{table}

{\bf Table 3.}
Displacements of U-shaped trajectories with $x_0=18$ m and different
orbital altitudes for $\delta J_2=10^{-4}$.  The second line shows the
values of $\Horb$.

\begin{center}
\begin{tabular}{|r|r|r|r|}
\hline
    $y_0$   & \multicolumn{3}{|c|}
                { \wide $\delta x_{\max} \times 10^{7}$ (m)}  \\
\cline{2-4}
   (cm)     & 500 km  & 1500 km & 3000 km  \\
\hline
	-25 & 42.6\z  & 13.3\z  & 4.84   \\
 	-20 & 11.9\z  & 5.51    & 2.18   \\
 	-15 & 4.02    & 1.98    & 0.86  \\
 	-10 & 1.03    & 0.56    & 0.28  \\
 	 -5 & 0.17    & 0.11    & 0.07  \\
\hline
\end{tabular}
\end{center}
\end{table}

The above results show that the Shepherd quad\-ru\-pole moment uncertainty
$\delta J_2$ may be neglected in the SEE experiment with circular Shepherd
orbits at $\Horb=1500$ or $3000$ km and U-shaped Particle trajectories if
$\delta J_2\leq 10^{-5}$ and the position measurement error $\delta l$ is
$10^{-6}$ cm, or $\delta J_2\leq 10^{-7}$ for $\delta l = 10^{-8}$ cm. For
cycloidal Particle trajectories or elliptic Shepherd orbits these estimates
become $\delta J_2\leq 10^{-4}$ and $\delta J_2\leq 10^{-6}$, respectively.
For low orbits, $\Horb=500$ km, the resitrictions on $\delta J_2$ become
more stringent:  $\delta J_2\leq 10^{-6}$ for $\delta l = 10^{-6}$ cm and
$\delta J_2\leq 10^{-8}$ for $\delta l = 10^{-8}$ cm. However, as is evident
from the above tables, these requirements may be relaxed by an order of
magnitude if one discards some trajectories.

The influence of $\delta J_2$ on the accuracy of $G$ measurement may be now
estimated as follows. Let some value of $\delta J_2$
produce the trajectory displacement $| \delta \vec{r}|
\leq \delta l_j$ while the variation $\delta G_0$ of $G$ with the same
initial conditions gives the trajectory displacement
$|\delta\vec {r}| \leq \delta l_{G}$. Then, keeping in mind the
linear dependence of trajectory displacements on $\delta J_2$ and
$\delta G$, the accuracy of $G$ measurement under the
uncertainty $\delta J_2$ may be estimated as
\[
\frac{\delta G}G\leq \frac{\delta l_j}{\delta l_{G}}\frac{\delta G_0}G.
\]

Using this inequality and the results of trajectory simulations,
we obtain the following estimates for U-shaped Particle trajectories
in circular orbits with $\Horb=1500$ km:

\medskip\noi
{\bf Table 4. } Estimates of $\delta G/G$ in ppm
for $\delta J_2=10^{-4}$, when the symmetry axis of the Shepherd lies in
($\chi=0$) or is ortogonal to ($\chi=\pi/2$) its orbital plane.
The second line shows $x_0$.

\begin{center}
\begin{tabular}{|r|r|r|r|r|}
\hline
$y_0,$ &
      \multicolumn{2}{|c|}{$\chi=0$}&
      \multicolumn{2}{|c|}{$\chi=\pi/2$} \\
\cline{2-5}
 cm  & 18 m & 6 m & 18 m & 6 m \\
\hline
	-25 & $0.88$ & $0.7$  &  $3.57$ & $6.7\z$  \\
\hline
	-20 & $0.27$ & $0.3$  &  $1.05$ & $1.73$  \\
\hline
	-15 & $0.06$ & $0.1$  &  $0.24$ & $0.44$ \\
\hline
\end{tabular}
\end{center}

One can conclude that the uncertainties $\delta J_2 \lsim 10^{-5}$
do not create substantial $G$ errors for most of the trajectories.

\section {Simulations of experimental procedures}

This section describes the results of computer simulations of the whole
measurement procedures aimed at obtaining the sought-after gravitational
interaction parameters. These simulations assumed the
Shepherd mass $M=500$ kg, a circular orbit with $\Horb=1500$ km under a
spherical gravitational potential of the Earth, and a Particle mass of $100$
g. Where relevant, it is assumed that both the Shepherd and the Particle are
made of tungsten. Their identical compositions are assumed for simplicity
since this work is performed only for estimation purposes.

\subsection{Equations of motion with Yukawa terms}

We will begin with a presentation of the Particle equations of motion in the
relevant approximation, including the contributions from hypothetical Yukawa
forces, taking into account the finite size of the Yukawa field sources.

     Let the interaction potential for two elementary masses $m_1$ and $m_2$
     be described by the potential
\beq                                                         \label{Yuk1}
     dV^{\rm Yu} = \frac{G\,dm_1 dm_2}{r} \alpha\e^{-r/\lambda}
\eeq
     where $r$ is the masses' separation, $\alpha$ and $\lambda$ are the
     strength parameter and the range of the Yukawa forces. Then for two
     massive bodies with the radii $R_1$ and $R_2$
     after integration over their volumes we obtain \cite{ZaKol}
\beq
     V^{\rm Yu} = \frac{G\,m_1 m_2 \beta_1 \beta_2}{r}      \label{Yuk2}
                      \alpha\e^{-r/\lambda}
\eeq
     where
\beq                                                         \label{Yuk3}
     \beta_i = 3 \biggl(\frac{\lambda}{R_i}\biggr)^3
       \biggl[\frac{R_i}{\lambda}\cosh \frac{R_i}{\lambda}
	      		- \sinh \frac{R_i}{\lambda} \biggr].
\eeq
     When $R_i/\lambda \ll 1$, we have $\beta_i \approx 1$.
     This may be the case when we consider the interaction between the
	Shepherd and the Particle at a distance of the order of a few metres.
	The radii of the Shepherd and the Particle are small: $R_1 \approx
	18$ cm for the Shepherd and $R_2 \approx 1.1$ cm for the Particle.
	If the range $\lambda$  is of the order of the Earth radius, $\lambda
	\approx \RE$, we have $\beta_{\oplus}=1.10$ and $\beta_{1,2}=1$ where
	the indices 1 and 2 label the Shepherd and the Particle, respectively.

     The equations of motion are obtained under
     the following assumptions. There are two Yukawa interactions
     with the parameters $\lambda_0$ and $\alpha_0$
     referring to the Earth-Shepherd and Earth-Particle interactions which
     are the same (due to the assumed identical composition for the Shepherd
     and the Particle), while $\lambda$ and $\alpha$
     determine the Shepherd-Particle interaction.
     The equations of motion in the frame of reference connected with the
     Shepherd, with the same notations $x$, $y$, $s$ as previously, are
\bearr
     \ddot{x} +2\omega^2 \dot{y}
      		+ G(m_1+m_2) \frac{x}{s^3} - 3\omega^2 \frac{xy}{s} \nnn
\cm     + G(m_1+m_2) \frac{x}{s^3}
        	 \alpha \biggl(1+ \frac{s}{\lambda}\biggr)\e^{-s/\lambda}
	=0;  						              \nnnv
     \ddot{y}
     -2 \omega\dot{x} - 3\omega^2 y + G(m_1+m_2)\frac{y}{s^3}
	    + \frac{3\omega^2}{r_{01}} \biggl(y^2 -\frac{x^2}{2}\biggr)\nnn
\cm     +G(m_1+m_2)\frac{y}{s^3}
	    \alpha \biggl(1+\frac{s}{\lambda}\biggr)\e^{-s/\lambda} \nnn
\inch\ \ \ - \omega^2 \beta_0 \alpha_0 \e^{-r_{01}/\lambda_0}y =0
							      \label{Yuk4}
\ear
      where $\omega$ is the orbital frequency:
\beq                                                          \label{omega}
	\omega^2 = \frac{G\ME}{r_{01}^3}
	\biggl[ 1+\beta_0 \alpha_0 \biggl(1+\frac{r_{01}}{\lambda_0}\biggr)
		\e^{-r_{01}/\lambda_0}\biggr].
\eeq
      We have neglected the terms quadratic in $s/r_{01}$ times $\alpha$ or
      $\alpha_0$ due to their manifestly small contributions.

      If we set $\alpha_0=0$ in \eqs (\ref{Yuk4}), we obtain the equations
      used to describe only the Shepherd-Particle Yukawa interaction.
      One can notice that Yukawa terms are roughly proportional to the
      gradients of the corresponding Newtonian accelerations, namely,
      $Gm_1/s^3$ for the Shepherd-Particle interaction and
      $G\ME/r_{01}^3 \approx \omega^2$ for (say) the Earth-Shepherd
      interaction. In our case these quantities are estimated as
\bearr
      \frac{Gm_1}{s^3} \approx 2.7\ten{-10}\ {\rm s}^{-2}
		       \cm {\rm for}\ \ s=5\ {\rm m},\nnn \cm
      \omega^2 \approx 8.16\ten{-7} \ {\rm s}^{-2}.
\ear
     Thus, given the same strength parameter, the Earth's Yukawa force is
     three orders of magnitude greater than that between the Shepherd and
     the Particle, therefore one might expect some significant progress in
     an ISL test for $\lambda$ of the order of the Earth's radius.

     The effect of the Earth's Yukawa force is proportional to the
	displacements of the statellite along the direction of the Earth's
	radius. Therefore the sensitivity of the SEE method will increase
	if one uses orbits with eccentricities of the order of 0.01,
	following Nordtvedt's suggestion \cite{Nor}. (Larger eccentricities
	would too much disturb the qualitative picture of a SEE encounter.)
	Tentative estimates show that in this way one can achieve sensitivities
	to $\alpha \sim 10^{-10}$, and more thourough studies are in progress.

     \eqs (\ref{Yuk4}) were used to simulate the measurement procedures.

\subsection{Simulations of an experiment for measuring $G$}

     The constant $G$ is determined from the best
     fitting condition between the ``theoretical''
    ($\rt (t_i) = \rt_i$) and ``empirical'' ($\vec r_i$) Particle
    trajectories near the Shepherd. The fitting quality is evaluated by
    minimizing a functional characterizing a ``distance'' between the
    trajectories. We have considered the following functionals for such
    ``distances'':
\bearr
     S = \sum_{i=1}^{N}
    \biggl [(x_i-\xt_i)^2 + (y_i-\yt_i)^2 \biggr], \label{F1}\\
\lal     S_x = \sum_{i=1}^{N}
     			(x_i-\xt_i)^2, \qquad
       S_y = \sum_{i=1}^{N}
     	                (y_i-\yt_i)^2,         \label{F2}    \\
\lal     S^* = \sum_{i=1}^{N}
    \biggl [|x_i-\xt_i| + |y_i-\yt_i| \biggr], \label{F3} \\
\lal     S_x^* = \sum_{i=1}^{N}
     	           |x_i-\xt_i|, \qquad
       S_y^* = \sum_{i=1}^{N}
     	           |y_i-\yt_i|. \label{F4}
\ear
The theoretical trajectory depends on the gravitational constant $G$, on the
initial coordinates $x_0, y_0$ and on the initial velocities $v_{x0},
v_{y0}$.  To estimate $G$, one chooses the value for which a ``distance''
functional in the space of the five variables ($G, x_0, y_0, v_{x0}, v_{y0}$)
reaches its minimum.


\Picture{0}{55}{\special{em:graph pic1.gif}}
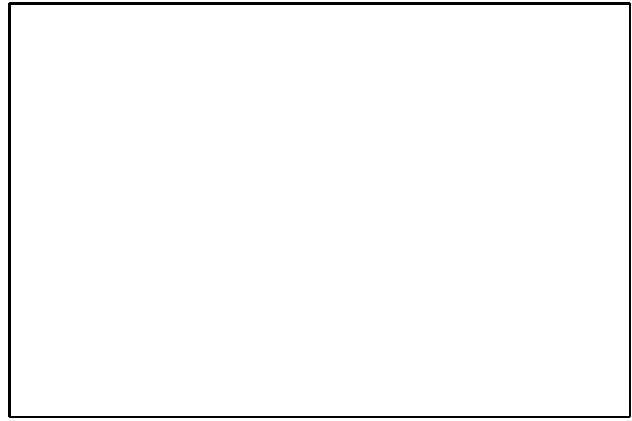
     {Errors $\delta G$ estimated
	by the gradient descent ($R_{\rm grad}$) and consecutive descent
	($R_{\rm s}$) methods}

We carried out a computer simulation of the SEE experiment and estimated
$\delta G$ for a given coordinate measurement error
($\sigma=1\ten{-6}$ m). As ``empirical'' trajectories, we took computed
trajectories, with specified values of the above five variables, where a
Gaussian noise was introduced from a random number generator. Independent
``empirical trajectories were created by non-intersecting random number
sequences. The functional was minimized using the gradient
descent method and the consecutive descent method.
The starting value of the ``vertical'' (along the Earth's radius)
coordinate, $y_0$, was taken to be 0.25 m, while the horizontal one, $x_0$,
varied between 2 and 18 m. Fig\,1 
shows the dependence of the errors
$\delta G/G = R_{\rm grad}$, obtained by
the gradient descent method and $\delta G/G = R_{\rm s}$, obtained by the
consecutive descent method.  All the errors are estimated by confidence
intervals corresponding to a confidence of 0.95. The mean values of these
errors are as follows:
\bear
	 R_{\rm grad} = 4.69\ten{-8},\cm
      R_{\rm s} \eql 5.24\ten{-8}.
\earn
Thus the errors estimated by the gradient and consecutive descent methods
are close to each other and are about an order of magnitude smaller than the
error from one-trajectory data. It has been discovered that the simulation
results strongly depend on the random number generator, so that ordinary
generators are not perfect.

The use of truncated functionals like (2) has shown that a functional
incorporating the more informative ``horizontal'' coordinate $x$ leads to
estimates close to those obtained from the total functional, whereas the use
of $y$ alone substantially decreases the sensitivity. Therefore in
practice, to determine $G$, it is sufficient to measure only one of the two
coordinates, viz. $x$.

Since the ``empirical'' trajectory is built on the basis of a computed one,
with a known value of the gravitational constant $G_0$, it appears
possible to estimate a possible systematic error inherent in the data
processing method. The latter has turned out to be in most cases much
smaller than the random error. This result shows the correctness of the
methods used.

As is evident from the results, the best accuracy is achieved at values of
$x_0$ ($\approx$ the capsule size) about 4--5 metres.

\Picture{0}{68}{\special{em:graph pic2.gif}}
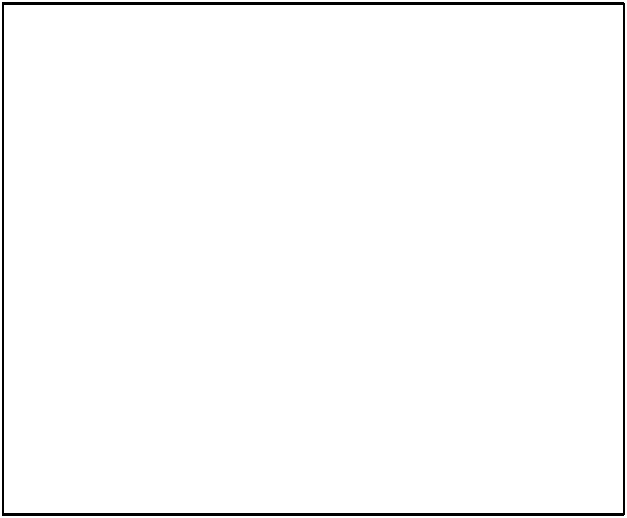
{The SEE method sensitivity to Yukawa forces between
the Shepherd and the Particle}

\subsection{\nhq Sensitivity to Yukawa forces with $\lambda \sim 1$\,m}

In an experiment for finding a Yukawa interaction between the
Shepherd and the Particle with the potential (\ref{Yuk2})
with $\beta_{1,2}=1$, one computes
two theoretical trajectories: one ignoring the
Yukawa forces ($x^0(t_i),\ y^0(t_i)$) and another taking
them into account $\Bigl(x^\alpha(t_i),\ y^\alpha(t_i)\Bigr)$. These two
computed curves are compared with the empirical trajectory
using the functional
$S_k$ ($k= 0, \alpha$) according to (\ref{F1}) which may be considered as a
dispersion characterizing a scatter of the ``empirical'' coordinates with
respect to the fitting trajectory. This is true when the theoretical
model is adequate to the real situation. In the case $k=\alpha$ the
functional $S_k= s_\alpha$ has a $\chi^2$ distribution with $n_2=2N-1$
degrees of freedom. With $k=0$ the parameter $\alpha$ is absent,
therefore $S_0$ is distributed according to the $\chi^2$ law with $N_1=2N$
degrees of freedom. Then their ratio $S_0/S_\alpha = F_{n_2,n_1}$ will be
distributed according to the Fischer law with $n_2$ and $n_1$ degrees of
freedom. If an experiment shows that, on a given significance level $q$,
the relation
\beq
	S_0/S_\alpha \geq F_{n_1,n_2, q}                       \label{F5}
\eeq
is valid, one should conclude that a Yukawa force has been
detected.  An equality sign shows a minimum detectable force on the given
significance level $q$. We have assumed $q=0.95$.
The results of a sensitivity computation for different
values of the space parameter $\lambda$ are presented in Fig.\,2.
A maximum sensitivity of $\alpha = 2.1\ten{-7}$ has been observed for
$\lambda= 1.25$ m.  This value is 3 to 4 orders of magnitude better than the
sensitivity of terrestrial experiments in the same range.

\Picture{0}{68}{\special{em:graph pic3.gif}}
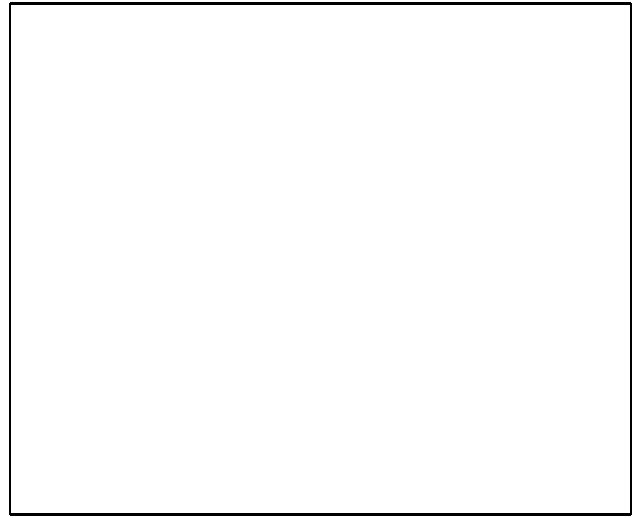
{The SEE method sensitivity to Yukawa forces with the range parameter
$\lambda_0$ of the order of the Earth's radius $\RE$}

     These results are based on the measurement method which was proposed in
     the original SEE paper \cite{SD92}; as already mentioned, a method
     involving an eccentric orbit \cite{Nor}, is much more sensitive and, by
     our tentative estimates, can give an error
     $\delta\alpha \lsim 10^{-10}$.

\subsection{Sensitivity to Yukawa forces with $\lambda \sim \RE$}

     To estimate the parameter $\alpha_0$ in \eqs(\ref{Yuk4}),
     computer simulations were carried out using the method
     as described above for $\alpha$, based on the Fischer
     criterion for the significance level 0.95. The range parameter
     $\lambda_0$ varied from $(1/32)\RE$ to $32\RE$.
     Two trajectories with the initial
     Shepherd-Particle separations $x_0$ of 2 and 5 m were calculated. In
     both cases the impact parameter $y_0$ was chosen to be 0.25 m. We used
     \eqs (\ref{Yuk4}) with $\alpha=0$, i.e., excluding the non-Newtonian
     interaction between the Shepherd and the Particle.
     As is evident from \eqs (\ref{Yuk4}), the Particle
     trajectory depends on the ratio $r_{01}/\lambda_0$ in the product
     $(r_{01}/\lambda_0)\e^{-r_{01}/\lambda_0}$.
     This quantity reaches its maximum at $\lambda_0=r_{01}/2$.
     Our calculations have confirmed that a maximum
     sensitivity of the SEE method ($3.4\ten {-8}$ for $x_0=5$ m)
     is indeed observed at this value of $\lambda_0$. This is about an order
     of magnitude better than the estimates obtained by other methods.
     Hopefully this estimate may be further improved by about an order of
     magnitude by optimisation of the orbital parameters.
     However, there is a factor which can, to a certain extent, spoil these
     results, namely, the uncertainty in the parameter $\omega$ which,
     in this calculation, was assumed to be precisely known.

     The simulation results are shown in Fig.\, 3  
     for two trajectories with initial Shepherd-Particle separations of 2
     and 5 metres.

\section
{A possible effect of the Earth's radiation belt}


Charged particles, penetrating into the SEE capsule from space
and captured by the test bodies, create electrostatic forces that could
substantially distort the experimental results. Among the sources of such
particles one should mention (i) cosmic-ray showers, (ii) solar flares and
(iii) the Earth's radiation belts (Van Allen belts). The effect of
cosmic-ray showers was estimated in Ref. \cite{SD92} and shown to be
negligible.  Solar flares are more or less rare events and, although they
create very significant charged particle fluxes, sometimes even exceeding
those in the most dense regions of the radiation belts, one can assume that
the SEE measurements (except those of $\dot G$) are stopped for the period
of an intense flare.  On the contrary, the effect of the Van Allen belts is
permanent as long as the satellite orbit passes, at least partially, inside
them.

We will show here that the charging is unacceptably high at otherwise
favourable satellite orbits, so that some kind of charge removal technique
is necessary, but this problem may be solved rather easily by
presently available technology.

The range of the most favourable SEE orbital altitudes, roughly 1400 to 3300
km \cite{SD92}, coincides with the inner region of the so-called inner
radiation belt \cite{2}--\cite{5}, situated presumably near the plane
of the magnetic equator.  This region is characterized by a
considerable flux of high-energy protons and electrons.
For a SEE satellite at altitudes near 1500 km the duration of the charging
periods is about 12 minutes.  Maximum charging rates occur in the central
Atlantic. It should be noted that the South Atlantic Anomaly (SAA) ---
a region of intense Van Allen activity which results from the low
altitude of the Earth's magnetic field lines over the South Atlantic
Ocean --- cannot cause additional problems for the SEE experiments.
The reason is that the SAA mostly contains low-energy protons which cannot
penetrate into the SEE capsule.

Electrons are known to be stopped by even a thin metallic shell, so
only protons are able to induce charges on the test bodies.
Proton-induced charges on the test bodies can create considerable forces.
The inner radiation belt contains protons with energies of 20 to
800 MeV, and their maximum fluxes at an altitude of 3000 km over the equator
are as great as about $3\ten 6 \flun$ for energies
$E\gsim 10^6$ eV and about $ 2\ten 4 \flun$ for $E\gsim 10^7$ eV.
At 1500 km altitude these numbers are a few times smaller; the
fluxes gradually decrease with growing latitude $\varphi$ and actually
vanish at $\varphi\sim 40\deg$.

It is thus necessary to have some estimates taking into account
that (i) the capsule walls have a considerable thickness and stop the
low-energy part of the proton flux and (ii) among the protons that penetrate
the capsule and hit the Particle, the most energetic ones, whose path in the
Particle material is longer than the Particle diameter, fly it through and
hit the capsule wall again. As for the Shepherd, its size is large enough to
stop the overwhelming majority of protons which hit it.

In what follows, we will assume a Shepherd radius of 20 cm and a Particle
radius of 2 cm and estimate the captured charges for some satellite orbits
in a capsule whose walls of aluminium are 2, 4, 6 and 8 cm thick.
The SEE satellite must actually involve several coaxial cylinders for
thermal-radiation control, and the combined thickness of their walls
must amount to several cm.
We will assume, in addition, that the Particle also consists of aluminium
and stops all protons whose path is shorter than 4 cm (thus a little
overestimating the charge since most of protons will cover a smaller
path through the Particle material). A 100 g Particle of aluminium will have
a radius of $\approx 2.07$ cm.

It is advisable to determine first which charges (and fluxes that create
them) might be regarded negligible.

\subsection {Admissible charges}

    Let us estimate the Coulomb interaction both between the Shepherd and
    the Particle and between each test body and its image in the capsule
    walls. To estimate the spurious effects on the
    Particle trajectory, it is reasonable to calculate its possible
    displacements due to the Coulomb forces from the growing captured
    charges. We assume that the test bodies are discharged by grounding
    to the capsule before launching the motion.

\medskip\noi
    {\bf Criterion.} We will call the induced charges, or the fields they
    create, {\sl admissible} if they cause a displacement of the Particle
    with respect to the Shepherd smaller than a prescribed coordinate
    measurement error $\delta l$ (we take here $\delta l = 10^{-6}$ m) for
    a prescribed measurement time (we take $t \geq 10^4$ s).

\medskip
    A charge on the Shepherd can be estimated as
\beq
     q_M \approx eS_M \int J(t,x)\,dt  = e S_M F(t,x)       \label{B2}
\eeq
    where $e$ is the elementary charge,
    $x$ is the capsule wall thickness in cm; $J(t,x)$
    is the integral proton flux in \flun after passing through the wall,
    that is, the flux of protons with energies $E_p > E_p(x)$ where $E_p(x)$
    is such an energy that the proton path in aluminium equals $x$ cm; $S_M
    \approx 1256$ cm$^2$ is the Shepherd's cross-section; $F(t,x)$ is the
    fluence, i.e., the total number of protons of relevant energies that
    crosses a square centimeter of area for a certain period $t$.

    In a similar way, the charge captured by the Particle may be found as
\bear
    q_m \al\lsim\al eS_m\int [J(t,x)-J(t,\,x{+}4)]\,dt    \label{B3} \nn
	   \eql e S_m \ [F(t,x)-F(t,\,x{+}4)]
\ear
    where $S_m \approx 12.56$ cm$^2$ is the Particle cross-section.
    The subtraction in the square brackets takes into account the protons
    which fly through the Particle without stopping there. The sign
    $\lsim$ is used since the effective Particle cross-section is smaller
    than its equatorial section.

    The Coulomb acceleration $a_Q (t) = q_M q_m /(r^2 m)$
    (in the Gaussian system of units) depends on the Shepherd-Particle
    separation $r$ and on the form of the function $J(t)$, which in turn
    depends on the satellite orbital motion.

    The charge-induced Particle displacement is approximately
\beq
	\Delta l = \int dt \biggl[\int dt\,a_Q(t)\biggr]
\eeq
    since the acceleration is almost unidirectional. If, for estimation
    purposes, we suppose that the flux is time-independent, $J=J_0=$const,
    and take into account that in \eq (\ref{B3}) the difference
    $J(t,{x})-J(t,x+4) \approx {2 \over 5}J(t,x)$
    (or even smaller; see particular values in the next section), then the
    resulting displacement is about
\beq
    \Delta l \sim \frac{1}{30}\frac{e^2 S_M S_m J_0^2 t^4}{r^2 m}.\label{B5}
\eeq
    The strong time dependence is explained by the
    rapid growth of the Coulomb force with capturing the charge. Numerically,
    with the above values of $S_M$ and $S_m$, taking $m=100$ g and $r=1$ m
    (the latter leads to an overestimated force since the Particle
    spends most of time at greater distances), one gets:
\beq
    J_0^2 t^4 \lsim 0.83\ten{18}\ {\rm s}^2 {\rm cm}^{-4}.     \label{B6}
\eeq
    For $t= 10^4$ s an admissible flux is only within 9 \flun.

    Another undesired effect is that the Particle, being charged by the belt
    protons, will interact with the capsule walls. This is well
    approximated as an interaction with the Particle's mirror image in the
    wall, while the latter may be roughly imagined as a conducting plane.
    Then, assuming that the Particle is at average at about 25 cm from the
    capsule wall and using the same kind of reasoning as above, we obtain
    instead of (\ref{B6})
\beq
	J_0^2 t^4 \lsim 2.07\ten{19}\ {\rm s}^2 {\rm cm}^{-4} \label{B7}
\eeq
    and an admissible proton flux within 45 \flun for $t=10^4$ s.

\begin{table*}          
\noi {\bf Table 5.}
     Average proton fluxes in some
     satellite orbits at minimum solar activity ($x$ is given in cm, $E_p$
     in MeV; $i$ is the orbit inclination and $\Omega$ is its ascension
     angle, i.e. the longitude at which the satellite crosses the equatorial
     plane moving northward. The notation 1.2(3) means $1.2\ten{3}$, etc.
     In the first column, the letter `a' labels equatorial orbits, `b' and
     `c' mark less and more favourable orbits (thst is, with greater and
	smaller numbver of protons), respectively, for given altitide and
	inclination.

\begin{center}
\begin{tabular}{|c|r|r|r|r|r|r|r|r|r|r|}
\hline
       & & & & \multicolumn{7}{|c|}{Integral flux for $E_p > E_p(x)$}\\
\cline{5-11}
 Orbit & $T$   &  $i$   &$\Omega$&$x=0$\z & $ x=2$\z& $x=4$\z & $x=6$\z  & $x=8$\z  & $x=10$\z & $x=12$\z \\
       &  s    &        &        &$E_p>0$ & $E_p>65$& $E_p>98$& $E_p>124$& $E_p>146$& $E_p>166$& $E_p>184$\\
\hline
 500b & 5677 &  89\deg & 23.7\deg  & 1200  & 8    & 4.5    &  3   &  2.1    &  1.8    &   1.5   \\
 800a & 6053 &   0\deg &           & 114   & 66   & 51\zz  &  41  &  33\zz  &  28\zz  &   24\zz \\
 800b & 6053 &  89\deg & 19.4\deg  & 2660  & 71   & 46\zz  &  34  &  26\zz  &  21\zz  &   17\zz \\
1000a & 6307 &   0\deg &           & 515   & 325  & 255\zz & 210  & 165\zz  & 140\zz  &  120\zz \\
1000b & 6307 &  89\deg & 20\deg\zz & 4120  & 173  & 116\zz &  89  &  65\zz  &  54\zz  &   44\zz \\
1000c & 6307 &  89\deg & -83\deg\zz& 51    & 26   & 20\zz  &  16  &  12\zz  &  10\zz  &   8.5   \\
1500a & 6960 &   0\deg &           & 6905  &3160  &2360\zz & 1850 & 1470\zz & 1260\zz & 1070\zz \\
1500b & 6960 & 102\deg & -12\deg\zz&1.2(4) &1420  &1000\zz & 770  & 600\zz  & 500\zz  &  415\zz \\
1500c & 6960 & 102\deg &  97\deg\zz& 2264  &646   &464\zz  & 365  & 280\zz  & 236\zz  &  196\zz \\
3000a & 9040 &   0\deg &           &2.6(5) &1.7(4)&1.15(4) & 9100 & 6500\zz & 5400\zz & 4350\zz \\
3000b & 9040 & 112\deg & -14.5\deg &1.6(5) &3700  &2400\zz & 1750 & 1300\zz & 1060\zz & 850\zz  \\
3000c & 9040 & 112\deg &  98\deg\zz&9.8(4) &3540  &2350\zz & 1700 & 1300\zz & 1060\zz & 850\zz  \\
\hline
\end{tabular}
\end{center}
\end{table*}

    Some more estimates are of interest: if a charge can be kept smaller
    than a certain value, then how great may it be to create only negligible
    displacements? Suppose that there are constant charges on both the
    Shepherd ($q=q_M$) and the Particle ($q=q_m,\ m=100$ g), then they are
    admissible according to the above criterion as long as
\bear
       q_M q_m \lal <2\ten{-6}\,\qg^2 = \fract{2}{9}\ten{-24}\ {\rm C}^2,
                                                               \label{B8}\\
         q_m^2 \lal <\half\ten{-6}\,\qg^2.                     \label{B9}
\ear
    These inequalities follow, respectively, from considering
    the Shepherd-Particle interaction and the
    interaction between the Particle (located at 25 cm from the
    wall) and its image. Thus the maximum admissible Particle charge is
    about $7\ten{-4}\,\qg \approx 1.5\ten{6} e$; assuming this value, it
    follows from (\ref{B8}) that the maximum Shepherd charge is about
    $3\ten{-3}\,\qg \approx  5.5 \ten{6} e$. With these
    charge values the electric potentials on the test body surfaces are
\bearr
    U_M \approx 1.5\ten{-4}\,\qg/{\rm cm} = 45\ {\rm mV}; \nnn
    U_m \approx
                3.5\ten{-4}\,\qg/{\rm cm} = 105\ {\rm mV}.    \label{b10}
\ear

    If by any means the requirements (\ref{B8}), (\ref{B9}) are satisfied
    (e.g., the potentials are kept smaller than the values (\ref{b10})),
    the electrostatic effect on the Particle trajectory may be neglected.

    The Shepherd's interaction with its image charge induced in its nearest
    bottom of the SEE experimental chamber does not lead to appreciable
    Particle displacements. A very demanding requirements on the
    Shepherd's charge emerges, however, if the SEE satellite is used for
    G-dot determination (whose detailed discussion is postponed to future
    papers). One obtains then
\beq
	U_M \lsim 12\ {\rm mV}.
\eeq
    Evidently, in this case the Shepherd-Particle interaction {\it per se\/}
    is not the determining factor with respect to charge limits on the test
    bodies.

\subsection{Evaluation of charges captured in some orbits}

    Let us now estimate the charges captured by the Shepherd
    and the Particle on board a satellite in various circular orbits for a
    single revolution around the Earth, a period of about two hours. Actual
    measurement times may exceed this period, but not too much.

    Approximate values of time-averaged proton fluxes are
    presented for some circular orbits in Table 5.

\begin{table*}
{\bf Table 6.}
Average flux, peak flux and captured charges per revolution in some
satellite orbits

{\small
\begin{center}
\begin{tabular}{|c|c|c|c|r|r|}
\hline
Orbit  &  Wall     &  Average flux,& Peak    flux,& Shepherd \z      &    Particle \z  \\
       & thickness &    \flun\z    &   \flun\z    & charge $q_M$     &    charge $q_m$ \\
\hline
       &  2 cm     &  \nhq 1420    & \nhq 12300  & 1.5\ten{10} $e$  &   4.5\ten 7 $e$ \\
1500b  &  4 cm     &  \nhq 1000    &      8800   &   1\ten{10} $e$  &   2.6\ten 7 $e$ \\
       &  6 cm     &   770         & 6800        & 7.5\ten 9 $e$    &   1.5\ten 7 $e$ \\
       &  8 cm     &   600         & 5400        &   6\ten 9 $e$    &   1.2\ten 7 $e$ \\
\hline
       &  2 cm     &   646         & 5700        & 6.5\ten 9 $e$    &   1.9\ten 7 $e$ \\
1500c  &  4 cm     &   464         & 4200        & 4.3\ten 9 $e$    &   1.1\ten 7 $e$ \\
       &  6 cm     &   365         & 3300        & 3.5\ten 9 $e$    &     7\ten 6 $e$ \\
       &  8 cm     &   280         & 2700        & 2.7\ten 9 $e$    &     5\ten 6 $e$ \\
\hline
\end{tabular}
\end{center}
}
\end{table*}

    The fluxes in Table 5 have been obtained using the computation
    software worked out at Nuclear Physics Institute (NPI) of Moscow State
    University, called SEE2 (Space Environment Effects 2) and SEREIS
    (Space Environment Radiation Effects Information System)
    \cite{7}--\cite{9}. This software made use of the NASA models AP8-max and
    AP8-min for calculating the proton fluxes \cite{10};
    however, the latter rest on measurements performed in
    the solar maximum of 1970 and minimum of 1964, while the NPI software
    uses some modern models of the Earth's magnetosphere, taking into
    account its evolution on the scale of decades.

    The high-energy particle fluxes in the radiation belts are
    strongly time-dependent; they vary between maxima and minima of solar
    activity, being, at least at low altitudes relevant for a SEE mission,
    greater at solar minima [2--5]. Table 5 shows the fluxes at a solar
    minimum; similar calculations for a solar maximum show {\it smaller\/}
    values by at average 20-25 per cent; the difference exceeds this value
    only for $x=0$ (being thus greater for low-energy protons than for
    higher-energy ones).

    The solar activity varies from one maximum or minimum to another,
    the Earth's magnetic field is sensitive to all these variations and also
    varies due to certain terrestrial phenomena. Another source of
    uncertainty, probably not a very strong one, is that the shielding effect
    is calculated by SEE2 software for a detector placed at the centre of a
    spherical shell of shielding material, whereas the capsule is
    cylindrical and the angular distribution of the proton flow is also
    uncertain. It is thus clear that any values like those presented in
    Table 5 may only serve as a guide, giving correct orders of magnitude.

    The above data make it possible to evaluate the captured charges. The
    results for two orbits, namely, 1500b and 1500c, are presented in Table
    6. These are the worst and the best variants of orbits at 1500 km among
    those analyzed (see the caption of Table 5).

    These and other similar data lead to some conclusions of importance
    for the SEE experiments.

    First, the models show zero proton fluxes in equatorial orbits of
    500 --- 800 km altitudes but indicate considerable fluxes at the same
    altitudes due to crossing the SAA. It turns out, however, that the SAA
    is overwhelmingly a low-energy phenomenon and almost does not affect
    fluxes on the relevant energy scale beginning with approximately 65 MeV.
    Even more, there is a very small proton flux due to the SAA even at
    energies over 10 MeV, so that behind a layer of 1 mm the SAA
    influence is already negligible. Therefore, behind a thicker metal layer
    there are actually no secondary particles due to SAA protons.

    Second, at 1500 km altitude the fluxes substantially depend on
    the orbit orientation but remain on the same scale of a few million
    protons per cm$^2$ at energies over 65 MeV.

    Third, evidently, at 3000 km altitude both the total flux (for $x=0$)
    and especially its high-energy part are a few times greater than at 1500
    km.

    Fourth, and most important: for all orbits in the desirable range of
    altitudes the charges are quite large as compared with their admissible
    values; they remain large even behind rather thick walls.
    It is thus quite necessary to have means to detect and remove the
    charges during the measurements.
    Moreover, as seen from the peak values in Table 6 and from time
    scans of Van Allen charging in orbits of interest (also obtained using
    the above-mantioned software), at a charging peak when crossing the
    magnetic equator the time required for the charge on the test bodies
    to reach its maximum allowable values, as listed above, is a matter of
    seconds, not minutes. Therefore the charge must be detected and removed
    as it builds up, on a time scale of seconds.

    The detection and measurement of the charge on the test
    bodies can probably be achieved relatively easily by an array of minute
    microvoltmeters attached to the inner wall of the experimental chamber.

    Several methods for removing positive charge are now being evaluated.
    A simple and promising method may be to shoot electron beams directly
    at test bodies. The number of electrons needed is on the order of
    $10^8/$sec.  Although this approach has the inherent drawback that it
    requires that an active system must perform correctly for many years, it
    is simple in principle and will accomplish the goal.

\section{Conclusion}

Space offers the prospect of quantum leaps in the accuracy of
gravitational experiments.  Although space is a challenging environment
for research, the inherent quiet of space can be exploited to make very
accurate determinations of $G$ and other gravitational parameters,
providing that care is taken to understand the many physical phenomena
in space which have the potential to vitiate accuracy.  A distinctive
feature of a SEE mission is its capability to perform such determinations
simultaneously on multiple parameters, making it one of the most promising
proposals.

To conclude, we enumerate different SEE tests and measurements
and show their expected accuracy as currently estimated:

\medskip\noi\
\begin{tabular}{ll}
{\it Test/measurement} \cm&    {\it Expected accuracy}
										\\[8pt]
EP/ISL at a few metres  &       $2\ten{-7}$               \yy
EP/CD at a few metres   &       $<10^{-7}\ (\alpha < 10^{-4})$ \yy
EP/ISL at $\sim\RE$     &       $<10^{-10}$                    \yy
EP/CD at $\sim\RE$      &       $<10^{-16}\ (\alpha <10^{-13})$\yy
$G$                     &       $3.3\ten{-7}$                 \yy
$\dot G/G$              &       $<10^{-13}$ in one year
\end{tabular}
\medskip

The last estimate is only tentative; the subject is under study.

\Acknow
{This work was supported in part by NASA grant \# NAG 8-1442.
K.A.B. wishes to thank Nikolai V. Kuznetsov for helpful discussions and for
providing an access to the SEE2 and SEREIS software.
}


\small
\begin{thebibliography}{99}

\bibitem{SD92}
    A.J. Sanders and W.E. Deeds, Phys. Rev. {\bf D 46}, 480 (1992).

\bibitem{dSMP}
    V. de Sabbata, V.N. Melnikov and P.I. Pronin,
    {\it Progr. Theor. Phys. \bf 88}, 623 (1992).

\bibitem{Mel94}
    V.N. Melnikov,
    {\it Int. J. Theor. Phys.\/} {\bf 33}, 7, 1569 (1994).

\bibitem{iztech}      
    A.D. Alexeev, K.A. Bronnikov, N.I. Kolosnitsyn, M.Yu.
    Konstantinov, V.N. Melnikov and A.G. Radynov,
    {\it Izmeritel'naya Tekhnika, } 1993, No. 8, 6; No. 9, 3; No. 10, 6;
    1994, No. 1, 3; {\it Int. J. Mod. Phys. }  {\bf D 3}, 4, 773 (1994).

\bibitem{Gi97}
    G.T. Gillies, {\it Rep. Progr. Phys.\/} {\bf 60}, 151 (1997).

\bibitem{LuTh82}
    G.G. Luther and W.R. Towler,
    {\it Phys. Rev. Lett.\/} {\bf 48}, 121 (1982).

\bibitem{FiArm}
    M.P. Fitzgerald and T.R. Armstrong,
    ``Recent Results for $g$ from MSL Torsion Balance'',
    {\it in:\/} ``The Gravitational Constant: Theory and Experiment 200
    Years after Cavendish'' (23-24 Nov. 1998), Abstracts
    (Cavendish--200 Abstracts), Inst. of Physics, London, 1998.

\bibitem{Nolting}
    J. Schurr, F. Nolting and W. K\"undig,
    {\it Phys. Rev. Lett.\/} {\bf 80}, 1142 (1998).


\bibitem{Meyer}
    H. Meyer,
    ``Value for $G$ Using Simple Pendulum Suspensions'',
    {\it in:\/} Cavendish--200 Abstracts.

\bibitem{Karag}
    O.V. Karagioz, V.P. Izmaylov and G.T. Gillies,
    {\it Grav. \& Cosmol.\/} {\bf 4}, 3, 239 (1998).

\bibitem{Richman}
    S.J. Richman, T.J. Quinn, C.C. Speake and R.S. Davis,
    ``Determination of $G$ Using the BIPM Torsion Balance'',
    {\it in:\/} Cavendish--200 Abstracts.

\bibitem{Schwarz}
    J.P. Schwarz, J.E. Faller, D.S. Robertson and T.M. Niebauer,
    ``The Free-Fall Determination of $G$,
    {\it in:\/} Cavendish--200 Abstracts.

\bibitem{PTB}
    W. Michaelis,
    ``A New Determination of the Gravitational Constant",
    {\it Bull. Am. Phys. Soc. \bf 40}, 2, 976 (1995).

\bibitem{Fis86}
    E. Fischbach and C. Talmage,
    {\it Mod. Phys. Lett.\/} {\bf A4}, 2303 (1989);
    {\it Nature (London)\/} {\bf 356}, 207 (1992).

\bibitem{Roll64}
    P.G. Roll, R. Krotkov and R.H. Dicke,
    {\it Ann. Phys. (N.Y.)\/} {\bf 26}, 442 (1964).

\bibitem{Brag71}
    V.B. Braginsky and V.I. Panov, {\it ZhETF} {\bf 61}, 873 (1971)
    [{\it Sov. Phys. JETP\/} {\bf 34}, 463 (1972)].

\bibitem{Adel94}
    E.G. Adelberger,
    {\it Class. Quantum Grav. \bf 11}, A9--A21 (1994).

\bibitem{Fuj71-72}
    Y. Fujii, {\it Nature Phys. Sci.\/} {\bf 234}, 5 (1971);
    {\it Ann. Phys. (N.Y.)\/} {\bf 69}, 494 (1972).

\bibitem{Long76-84}
    D.R. Long, {\it Nature (London)\/} {\bf 356}, 417 (1976);
    {\it Nuovo Cim.\/} {\bf 55B}, 252 (1984).

\bibitem{Adel+}
    E.G. Adelberger et al.,
    {\it Phys. Rev. Lett. \bf 59}, 8, 849--852 (1987);
    {\it Phys. Rev. \bf D 42}, 10, 3267--3292 (1990).

\bibitem{Adel97}
    E.G. Adelberger, 1997, private communication.

\bibitem{Ach}
    V. Achilli et al.,
    {\it Nuovo Cim. \bf 112B}, 5, 775 (1997).

\bibitem{FiG}
    E. Fischbach, G.T. Gillies, D.E. Kraus, J.G. Schwan, and C. Talmadge,
    {\it Metrologia \bf 29}, 3, 13--260 (1992).

\bibitem{Franklin}
    A. Franklin, ``The Rise and Fall of the Fifth Force:  Discovery, Pursuit,
    and Justification in Modern Physics'', American Institute of Physics,
    N.Y., 1993.

\bibitem{STEP}
J.-P. Blaser et al.,
``STEP (Satellite Test of Equivalence Principle):  Report on the
Phase A Study'', ESA/NASA report SCI (93) 4 (March, 1993);
``STEP (Satellite Test of Equivalence Principle):
Assessment Study'', ESA report SCI (94) 5 (May, 1994).

\bibitem{ZaKol}
    N.A. Zaitsev and N.I. Kolosnitsyn. {\it In:\/} ``Experimental Tests of
    Gravitation Theory'', Moscow University Press, 1989, p.38--50
    (in Russian).

\bibitem{Nor}
    K. Nordtvedt, private communication, 1997.

\bibitem{2}
    B.A. Tverskoy, ``Dynamics of the Earth's Radiation Belts'', Moscow,
    Nauka, 1968 (in Russian).

\bibitem{3}
    Yu.I. Galperin, L.S. Gorn and B.I. Khazanov, ``Radiation Measurments in
    Space'', Moscow, Atomizdat, 1972 (in Russian).

\bibitem{4}
    A.J. Dessler and B.J. O'Brien, ``Penetrating Particle Radiation'',
    {\it in:\/} ``Satellite Environment Handbook'', 2nd edition,
    ed. F.S. Johnson, Stanford University Press, Stanford, CA, 1965, p.
    53--92.

\bibitem{5}
    W.N. Hess, ``The Radiation Belt and Magnetosphere'', Ginn Blaisdell,
    Waltham, MA, 1968.

\bibitem{7}
    V.F. Bashkirov and M.I. Panasiuk,
    ``Dynamics of the Trapped Radiation Environment at Low Altitude Region of
    the Earth's Magnetosphere".
    Workshop ``Space Radiation Environment Modelling: New Phenomena and
    Approaches", Oct. 7--9, 1997: Program and Abstracts, p. 1.2.

\bibitem{8}
    V.F. Bashkirov, N.V. Kuznetsov and R.A. Nymmik,
    ``Information System for Evaluation of Space Radiation Environment and
    Radiation Effects on Spacecraft", ibid., p. 4.8.

\bibitem{9}
    V.F. Bashkirov and M.I. Panasiuk,
    {\it Kosmicheskiye Issledovaniya\/} (Space Research), {\bf 36}, 4,
    359--368 (1998).

\bibitem{10}
    D. Bilitza, ``Solar-Terrestrial Models and Application Software",
    {\it Planet. Space Sci.\/} {\bf 40}, 4, 541--579 (1992).

\end{thebibliography}
\end{document}